\documentstyle[amssymb,12pt]{article}

\begin{document}

\title{SOME HIGHER DIMENSIONAL VACUUM SOLUTIONS}
\author{ Metin G{\" u}rses\\
{\small Department of Mathematics, Faculty of Sciences}\\
{\small Bilkent University, 06533 Ankara - Turkey}\\
{\small and}\\
Atalay Karasu\\
{\small Department of Physics, Faculty of Arts and  Sciences}\\
{\small Middle East Technical University, 06531 Ankara-Turkey}}

\begin{titlepage}
\maketitle
\begin{abstract}

We study an even dimensional manifold with a pseudo-Riemannian metric
with arbitrary signature and arbitrary dimensions.
We consider the Ricci flat equations and
 present a procedure  to construct solutions to some higher (even)
dimensional Ricci flat field equations from the four dimensional Ricci flat
metrics. When the four dimensional Ricci flat geometry corresponds to
a colliding gravitational vacuum spacetime our approach provides
an exact solution to the vacuum Einstein field equations for colliding
gravitational plane waves in an (arbitrary) even dimensional spacetime. 
We give explicitly higher dimensional Szekeres metrics and study
their singularity behaviors.

\end{abstract}
\end{titlepage}

\section{Introduction}

In general theory of relativity there exist several solution
generating techniques for vacuum and electrovacuum Einstein field
equations \cite{krm}, \cite{dit}. These techniques basically give
constructions of metrics from the known metrics.
Recently \cite{gur1}, \cite{gur2} we have given a direct construction
of the metrics of the $2N$ dimensional Ricci flat geometries from the
two dimensional minimal surfaces in a pseudo Euclidean three geometry.
In this work we present a procedure to obtain solutions to some higher
dimensional Ricci flat field equations from some dimensional
Ricci flat metrics. We show that starting from
Ricci flat metric of a four dimensional geometry admitting two Killing
vector fields it is possible to generate a whole class $2N$
dimensional Ricci flat metrics. Here, in general, both the four
dimensional and $2N$ dimensional geometries have arbitrary signatures.
Among these there are some geometries have physical importance in
general theory of relativity and also in low energy limit of string
theory. For example, If the four dimensional geometry describes the 
colliding gravitational plane
wave geometry then the $2N$ dimensional geometry , for all $N>2$,
describes colliding vacuum  gravitational plane waves in higher
dimensional Einstein theory. We give direct construction of the 
$2N$ dimensional metrics from the four dimensional Ricci flat metrics.
As an explicit example we give a higher  dimensional extension
of the Szekeres \cite{szk} colliding vacuum gravitational plane wave metrics. 

The singularity structure of these higher dimensional solutions is
examined by using the curvature invariant. It is shown that the
singularity becomes weaker or stronger depending upon the parameters
of the solution.  Hence the singularity character of the solution may change
with the increasing number of dimensions.

Let $M$ be a $2N=2+2n$ dimensional manifold with a metric

\begin{eqnarray}
ds^2&=&g_{\alpha \beta}\,dx^{\alpha}\,dx^{\beta} \nonumber \\
&=&g_{ab}(x^{c})dx^a~dx^b+ H_{AB}(x^{c})dy^A~dy^B,   \label{a0}
\end{eqnarray}

\noindent
where $x^{\alpha}=(x^{a},\, y^{A})$ , $x^{a}$ denote the local coordinates 
on a 2-dimensional manifold and $y^{A}$ denote the local coordinates
on $2n$-dimensional manifold and $a,b=1,2$ ,$A,B=1,2,...,2n$.
The Christoffel symbols of the metric $g_{\alpha \beta}$ are given by

\begin{eqnarray}
\Gamma^{A}_{Ba}&=& {1 \over 2}H^{AD}~~H_{DB,a},~~~
\Gamma^{a}_{AB}=- {1 \over 2}g^{ab}~~H_{AB,b},
~~\bar \Gamma^{a}_{bc} =\Gamma^{a}_{bc},\\
\Gamma^{A}_{BD}&=& \Gamma^{A}_{ab}=  \Gamma^{a}_{AB}=0,
\end{eqnarray}

\noindent
where  the $\Gamma^{a}_{bc}$ are the Christoffel symbols of the
2-dimensional metric $g_{ab}$.

\noindent
The components of the Riemann tensor are given by

\begin{equation}
R^{\alpha}_{ \beta \gamma \sigma}=\Gamma^{\alpha}_{\beta \gamma,  \sigma}-
\Gamma^{\alpha}_{\beta \sigma, \gamma}+\Gamma^{\alpha}_{\rho \gamma}\,
\Gamma^{\rho}_{\beta \sigma}-\Gamma^{\alpha}_{\rho \sigma}\, 
\Gamma^{\rho}_{\beta \gamma}.
\end{equation}

\noindent
The components of the Ricci tensor are

\begin{eqnarray}
{\cal R}_{ab}&=&R^{\alpha}_{a \alpha b} \nonumber \\
&=&R_{ab}+{1 \over 4}tr(\partial_{a}H^{-1}\partial_{b}H)
-\bigtriangledown_{a} \bigtriangledown_{b}log \sqrt{\det{H}},\\
{\cal R}_{AB}&=&- {1 \over 2}(g^{ab}H_{AB,b})_{,a}
- {1 \over 2}g^{ab}H_{AB,b}[{(\sqrt{\det{g}})_{,a} \over \sqrt{\det{g}}}
+{(\sqrt{\det{H}})_{,a} \over \sqrt{\det{H}}}] \nonumber \\
&+&{1 \over 2}g^{ab}H_{EA,b}H^{ED}H_{DB,a}, \\
{\cal R}_{aA}&= &0,
\end{eqnarray}

\noindent
where $R_{ab}$ is the Ricci tensor of the 2-dimensional metric $g_{ab}$.

\section{Ricci flat geometries}

The Ricci flat conditions or the vacuum  Einstein field equations are
given by

\begin{equation}
\partial_{a}[\sqrt{\det{H}\,g}\;g^{ab}H^{-1}\partial_{b}H]=0,
\end{equation}

\begin{equation}
R_{ab}+ {1 \over 4}tr(\partial_{a}H^{-1}\partial_{b}H)-
\bigtriangledown_{a} \bigtriangledown_{b}log \sqrt{\det{H}}=0,
\end{equation}

\noindent
where $H$ is a $2nx2n$ matrix of $H_{AB}$ and $H^{-1}$ is its inverse
and $\bigtriangledown$ is the covariant differentiation with respect
to the connection $\Gamma^{a}_{bc}$ (or with respect to metric
$g_{ab}$). We may rewrite the 2-dimensional metric as

\begin{equation}
g_{ab}=e^{-M}\,\eta_{ab} ,\;\;\;
\end{equation}

\noindent
where $\eta$ is the metric of flat 2-geometry with arbitrary signature
($0$ or $\pm 2$ )  and the function $M$ depends on the local
coordinates $x^{a}$.
The corresponding Ricci tensor and the Christoffel symbol are

\begin{eqnarray}
R_{ab}&=& {1 \over 2} (\bigtriangledown_{\eta}^{2} M)\,\eta_{ab}, \nonumber \\
\Gamma^{c}_{ba}&=& {1 \over 2}[-M_{,b}\delta^{c}_{a}
-M_{,c}\delta^{a}_{b}+M_{,d}\eta^{cd}\eta_{ab}].
\end{eqnarray}

\noindent
Now let $H$ be a block diagonal matrix of $H_{AB}$ and each block is
a $2x2$ matrix

$$\pmatrix{\epsilon_{1}\,e^{u_{1}}h_{1} & & &\cr
&\ddots& &\bigcirc\cr
\bigcirc& &\ddots & \cr
& &  & \epsilon_{n}\,e^{u_{i}}h_{i} \cr}$$

\noindent
with $det {h_{i}}=1$  and $\epsilon_{i}=\pm 1$ for all $i=1,2, \cdots
, n$. Then

\begin{equation}
tr(\partial_{a}H^{-1}\partial_{b}H)=-2 \sum_{i=1}^{n}
 \partial_{a}u_{i}\partial_{b}u_{i}+ tr \sum_{i=1}^{n}
\partial_{a}h_{i}^{-1}\partial_{b}h_{i}
\end{equation}

\noindent
and

\begin{equation}
\det{H} =e^{2\,U} ,~~~ \sum_{i=1}^{n}\,u_{i}=U. \label{u01}
\end{equation}

\noindent
With the above anzats we can write the higher dimensional vacuum
field equations as

\begin{eqnarray}
{1 \over 2} \bigtriangledown_{\eta}^{2} M\eta_{ab}-U_{,ab}-
{1 \over 2}[M_{,a}U_{,b}+M_{,b}U_{,a}-M_{,d}U_{,}^{d}\eta_{ab}] \nonumber \\
-{1 \over 2}\sum_{i=1}^{n} \partial_{a}u_{i}\partial_{b}u_{i}
+ {1 \over 4} tr \sum_{i=1}^{n}\partial_{a}h_{i}^{-1}\partial_{b}h_{i}
=0  \label{a1}
\end{eqnarray}

\noindent
and

\begin{eqnarray}
\partial_{a}[\eta^{ab}e^{U}\partial_{b}u_{i}]=0, \label{aa1}\\
\partial_{a}[\eta^{ab}e^{U}h_{i}^{-1}\partial_{b}h_{i}]=0, \label{aa2}
\end{eqnarray}

\noindent
where there is no sum over $i$ (for all $i=1,2, \cdots n $).

\section{Four dimensional geometries}

We first consider the four dimensional case ($n=1$). We distinguish the
metric functions of the four dimensional case from the higher
dimensional ($n >1$) metric functions  by letting

\begin{equation}
M={\cal{M}},~~~ U={\cal{U}},~~~ h=h_{0}.
\end{equation}

\noindent
Since there are infinitely many possible solutions of the vacuum four
dimensional Ricci flat equations we shall denote 
${\cal{M}}_{i}, \,h_{0\,i}$ , $i=1,2, \cdots ,m$ to 
distinguish this difference. We label all these different solutions 
by putting a subscript $i=1,2, \cdots ,m$. Any two different solutions
either have different analytic forms or have the same analytic forms but 
with different integration constants. We assume that all these different
solutions have the same metric function $\cal{U}$. By this choice we
loose no generality because it is a matter of choosing a proper
coordinate system. The field equations are

\begin{eqnarray}
&&{1 \over 2} (\bigtriangledown_{\eta}^{2} {\cal{M}}_{i})\, \eta_{ab}-
{\cal{U}}_{,ab}-
{1 \over 2}[{\cal{M}}_{i,a}\, {\cal{U}}_{,b}+ {\cal{M}}_{i,b}\, 
{\cal{U}}_{,a}-{\cal{M}}_{i,d}\, {\cal{U}}_{,}^{d}\,\eta_{ab}]\nonumber
\\
&-&{1 \over 2} \partial_{a} {\cal{U}}\, \partial_{b}\,{\cal{U}}
+ {1 \over 4} tr (\partial_{a} \,h_{0\,i}^{-1}\partial_{b} \,h_{0\,i})
=0,\label{a2}
\end{eqnarray}

\noindent
and

\begin{eqnarray}
\partial_{a}[\eta^{ab}\,e^{{\cal{U}}} \partial_{b} {\cal{U}}]=0, \label{bb1}\\
\partial_{a}[\eta^{ab}\,e^{{\cal{U}}}\, h_{0\,i}^{-1} \partial_{b}
\,h_{0\,i}]=0. \label{bb2}
\end{eqnarray}

\noindent
For each $i=1,2, \cdots, m$ where $m$ is an arbitrary integer, each triple

$$({\cal{M}}_{i},\, h_{0\,i}, \,{\cal{U}})$$

\noindent
form a solution to the four dimensional vacuum field equations and we
assume that the function ${\cal{U}}$, for all these different
solutions, be the same.

\section{Higher dimensional Ricci flat geometries}

\noindent
We start with the assumptions that $U={\cal{U}}$ where the function $U$
is  defined in (\ref{u01}) , $h_{i}=h_{o\,i}$ and $m=n$ and using 
(\ref{a2}) into (\ref{a1}) we get

\begin{eqnarray}
&&{1 \over 2} \bigtriangledown_{\eta}^{2} (M-\tilde M)\eta_{ab}
+(n-1){\cal{U}}_{,ab} -{1 \over 2}[(M-\tilde M)_{,a}\, 
{\cal{U}}_{,b}\nonumber \\
&+&(M-\tilde M)_{,b}\, {\cal{U}}_{,a}-(M-\tilde M)_{,d}\, 
{\cal{U}}_{,}^{d}\, \eta_{ab}]
-{1 \over 2} \sum_{i=1}^{n}\partial_{a}\, u_{i}\partial_{b} u_{i} \nonumber\\
&+& {1 \over 2}n \partial_{a} {\cal{U}}\,\partial_{b}\, {\cal{U}}
=0,
\end{eqnarray}

\noindent
where $ \sum_{i=1}^{n} {\cal{M}}_{i}=\tilde M$.
Define $M-\tilde M =\bar M$, the above equation can be written as

\begin{eqnarray}
{1 \over 2} (\bigtriangledown_{\eta}^{2} \bar M)\, \eta_{ab}
+(n-1)\,{\cal{U}}_{,ab}-
{1 \over 2}[\bar M_{,a} \, {\cal{U}}_{,b} 
+\bar M_{,b}\, {\cal{U}}_{,a} \nonumber\\
-\bar M_{,d} \, {\cal{U}}_{,}^{d}\, \eta_{ab}]
-{1 \over 2} \sum_{i=1}^{n}\partial_{a}\,u_{i}\partial_{b}\,u_{i}
+ {1 \over 2}n \partial_{a}\, {\cal{U}}\partial_{b}\, {\cal{U}}
=0. \label{a3}
\end{eqnarray}

\noindent
We assume that ${\cal{U}}, h_{0 \, i}$ for $i=1,2, \cdots , n$ are
given functions of $x^{a}$.
Hence given ${\cal{U}}$ we can solve (\ref{aa1}) for $u_{i}$ with
$i=1,2, \cdots , n$, or

\begin{equation}
\nabla^{2}_{\eta}\, u_{i}+\eta^{ab}\, {\cal{U}}_{,a}\,u_{i,b}=0. \label{bb3}
\end{equation}

\noindent
Then inserting ${\cal{U}}, u_{i}$ and $h_{o\,i}$ in (\ref{a3}) we
solve the function $\bar M$. Then we have the following theorem.

\vspace{0.3cm}

\noindent
{\it Theorem: If ${\cal{U}}, \, h_{o\, i}$ and ${\cal{M}}_{i}$ , for
each $i=1,2, \cdots, n$, form a  solution to the four dimensional 
Ricci flat field equations for the metric

\begin{equation}
ds^2=e^{-{\cal{M}}_{i}}\, \eta_{a b}\,dx^{a}\,dx^{b}+
 e^{{\cal{U}}}\,(h_{0\,i})_{\,ab}\,dy^a~dy^b, ~~i=1,2,\cdots,n ,
\end{equation}

\noindent
where ${\cal {M}}_{i}={\cal {M}}_{i}(x^{a})$, ${\cal {U}}={\cal
{U}}(x^{a})$, and $h_{0\, i}=h_{0\, i}(x^{a})$, 
then the metric of $2n+2$ dimensional geometry defined below 

\begin{equation}
ds^2=e^{-M}\,\eta_{a b}\, dx^{a}\,dx^{b}+ 
\sum_{i=1}^{n}\, \epsilon_{i}\,e^{u_{i}}\,(h_{0\,i})_{ab}\,dy_{i}^a\,dy_{i}^b,
\end{equation}

\noindent
solves the Ricci flat equations, where $
\epsilon_{i}=\pm 1$, $M=\bar M+\tilde M $, $\tilde M=
\sum_{i=1}^{n}\, {\cal{M}}_{i}$ , $\bar M$ solves (\ref{a3}) and
$u_{i}$  solve (\ref{bb3}). Here
the local coordinates of the $2n+2$ dimensional geometry are given by
$x^{\alpha}=(x^{a}, y_{1}^{a}, y_{2}^{a}, \cdots, y_{n}^{a})$.
}

\vspace{0.3cm}

\noindent
We shall now consider some examples which will be obtained by the
application of the theorem. We shall consider the case which has a physical
importance as far as the Einstein's theory of general relativity is
concerned. We let $\epsilon_{i}=1$ for all $i=1,2, \cdots, n$ and 

$$\eta=\left(\begin{array}{cc} 0&1\\1&0 \end{array}\;
\right),~~~ x^{1}=u,\, x^{2}=v ,$$

\noindent
 then the equations in (\ref{a3}) become

\begin{eqnarray}
\partial_{u}\, {\bar M}\, \partial_{u}\,{\cal{U}}&=&
(n-1) \partial_{uu}\,{\cal{U}}-{1 \over 2}\, \sum_{i=1}^{n}\,
(\partial_{u}\, u_{i})^{2}+{n \over 2}\,
(\partial_{u}\,{\cal{U}})^{2}, \label{cc1} \\
\partial_{v}\,{\bar M}\, \partial_{v}\,{\cal{U}}&=&
(n-1)\, \partial_{vv}\,{\cal{U}}-{1 \over 2}\, \sum_{i=1}^{n}\,
(\partial_{v}\, u_{i})^{2}+{n \over 2}\,
(\partial_{v}\, {\cal{U}})^{2}, \label{cc2}
\end{eqnarray}

\noindent
where the $(uv)$ component of (\ref{a3}) is identically satisfied by
virtue of the equations (\ref{cc1}), (\ref{cc2}), (\ref{bb3}) and
(\ref{bb1}). The above equations remind us the construction
of the solutions of the  Einstein -Maxwell-massless scalar field
equations from the metrics of the Einstein- Maxwell spacetimes \cite{ers}. 
Eq.(\ref{bb3}) becomes

\begin{equation}
2\,u_{i,uv}+{\cal{U}}_{,u}\,u_{i,v}+{\cal{U}}_{,v}\,u_{i,u}=0. \label{cc3}
\end{equation}

\noindent
Hence for all $n>1$ to find a solution of higher dimensional colliding
gravitational vacuum plane waves   we have to solve the above
equations (\ref{cc1})-(\ref{cc3}) for ${\bar M}$ and $u_{i}~, i=1,2,
\cdots , n$. We shall now make a further assumption which solves
(\ref{cc3}) identically. Let $u_{i}=m_{i}\, {\cal{U}}$ where $m_{i}$,
($i=1,2, \cdots, n$) are real constants satisfying only the condition 

\begin{equation}
\sum_{i=1}^{n}\, m_{i}=1, \label{em}
\end{equation}

\noindent
otherwise they are arbitrary. Then the solution of (\ref{cc1}) and
(\ref{cc2}) can be found as

\begin{equation}
e^{-\bar M}=(f_{u}\,g_{v})^{-n+1}\,(f+g)^{{1 \over 2}\,(m^2+n-2)}.
\end{equation}

\noindent
Here we took 

\begin{equation}
e^{\cal{U}}=f(u)+g(v),
\end{equation}

\noindent
which is the general solution of (\ref{bb1}) where $f(u)$ and $g(v)$ are
arbitrary (differentiable) functions of $u$ and $v$ respectively and

\begin{equation}
\sum_{i=1}^{n}\,(m_{i})^2=m^2. 
\end{equation}

\noindent
Hence according to our theorem given above this completes the
construction of the metric of the corresponding vacuum spacetimes of 
dimension $2n+2$. Given any four dimensional metric of colliding
vacuum gravitational plane wave geometry (see \cite{gri} for this 
subject in detail) we have their extensions to higher dimensions for 
arbitrary $n$ without solving any further
differential equations. Sometimes to avoid some undesired
singularities on the whole $2n+2$
dimensional geometry it may be necessary to keep all the integration
constants of the original four dimensional metric variables 
$({\cal{M}}_{i}, {\cal{U}}, h_{0\,i})$. The boundary conditions
discussed
in \cite{szk} and in \cite{gri} (chapter 7, pages 46-47) of the four 
dimensional metrics should be used for the functions ${\cal{M}}_{i}$
to make them continuous across the boundaries $u=0,v=0$. Rather we
have to use them to make the $2n+2$ dimensional metric function $M$ 
to be continuous across these boundaries.

\section{Higher dimensional Szekeres solution}

For illustration let us take the Szekeres solutions \cite{szk}, \cite{gri}
(which contains the Khan-Penrose \cite{kp} solution as a special case)
 as the four dimensional vacuum solutions. 
They are given by

\begin{equation}
ds^2=2\,e^{-{\cal{M}}_{i}}\,du dv+e^{{\cal{U}}-V_{i}}\,dx^2+
e^{{\cal{U}}+V_{i}}\,dy^2, ~~i=1,2, \cdots, n
\end{equation}

\noindent
where 

\begin{eqnarray}
V_{i}=-2k_{i}\, \tanh^{-1}\, \bigl ({{{1 \over 2}-f} \over {{ 1 \over
2}+g}} \bigr )^{1 \over 2}-2 \ell_{i}\, \tanh^{-1}\, \bigl( {{{1 \over
2} -g} \over {{1  \over 2}+f}} \bigr)^{1 \over 2}, \label{vi}\\
{\cal{M}}_{i}=-\log(c_{i}f_{u}h_{v})-{1 \over  2}
(k_{i}^2+\ell_{i}^2+2k_{i} \ell_{i}-1)\, \log (f+g) \nonumber \\
+{k_{i}^2 \over 2}\, \log ({1 \over 2}-f)+{\ell_{i}^2 \over 2}\, \log ({1
  \over 2}-g)+{\ell_{i}^2 \over 2}\, \log ({1 \over 2}+f)+{k_{i}^2 \over
  2}\, \log ({1 \over 2}+g) \nonumber \\
+2k_{i}\, \ell_{i} \log \bigl (\sqrt{{1 \over 2}-f}\, \sqrt{{1 \over
    2}-g}+\sqrt{{1 \over 2}+f} \, \sqrt{{1 \over 2}+g} \bigr),
\end{eqnarray}

\noindent
where $k_{i}$, $\ell_{i}$, and $c_{i}$ are constants for all $i=1,2,
\cdots ,n$,  and 

\begin{equation}
f={1 \over 2}-(e_{1}\,u)^{n_{1}},~~~  g={1 \over 2}-(e_{2}\,v)^{n_{2}}. 
\end{equation}

\noindent
Here $e_{1}, e_{2}, n_{1} \ge 2$, and $n_{2} \ge
2$ are also
arbitrary constants. To avoid the discontinuity of the function
$e^{{-\cal{M}}_{i}}$ along the boundaries $u=0$ and $v=0$ some relations
among $k_{i}, \ell_{i}$ and $n_{1}, n_{2}$ are needed. We shall not set
these relations, because in our case the continuity of the function
$e^{-M}$ is important. For this purpose we give similar relations 
among these constants. Let us first define 

\begin{equation}
k^2=\sum_{i=1}^{n}\,k_{i}^2,~~~\ell^2=\sum_{i=1}^{n}\,
\ell_{i}^2,~~~s= \sum_{i=1}^{n}\,k_{i}\ell_{i},
\end{equation}

\noindent
and let

\begin{equation}
k^2=2(1-{1 \over n_{1}}),~~~\ell^2=2(1-{1 \over n_{2}}), \label{con}
\end{equation}

\noindent
where $n_{1} \ge 2, ~n_{2} \ge 2$.
We  observe that the constants $k$ and $\ell$ are
restricted to the range satisfying

$$ 1 \le k^2 < 2, ~~~ 1 \le \ell^2 <2.$$

\noindent
It is now easy to calculate $M$ which is continuous across the
boundaries $u=0$ and $v=0$ (by virtue of the conditions (\ref{con})). 
It reads

\begin{equation}
e^{-M}={(f+g)^{(k^2+\ell^2+m^2+2s-2) \over 2} \over ({1 \over 2}+f)^{{k^2 \over
2}}\,({1 \over 2}+g)^{{\ell^2 \over 2}}\, \bigl (\sqrt{{1 \over
2}-f}\,\sqrt{{1 \over 2}-g}+ \sqrt{{1 \over 2}+f}\, \sqrt{{1 \over
2}+g} \bigr)^{2s}}.
\end{equation}

\noindent
We also set $\Pi_{i=1}^{n}\,c_{i}=(e_{1}e_{2}n_{1}n_{2})^{-1}$.
Hence the metric of the $2n+2$ dimensional spacetime becomes

\begin{equation}
ds^2=2e^{-M}\,dudv+\sum_{i=1}^{n}\,(f+g)^{m_{i}}\,(e^{-V_{i}}\,dx_{i}^2+
e^{V_{i}}\,dy_{i}^{2}),
\end{equation}

\noindent
where $m_{i},~i=1,2,\cdots,n$ are constants with the condition given
in (\ref{em}) and $V_{i}$'s are given in (\ref{vi}). Here $x_{1}=x, y_{1}=y$.
When $n=1$ we have $m_{1}=1$, $m=1$ , $s=k \ell$ which corresponds to
the four dimensional case.

\section{Curvature singularities}

Next we calculate the curvature invariant of the metric (\ref{a0}).
The components of the Riemannian tensor are:

\begin{eqnarray}
R^{A}_{Bab}&=&\Gamma^{A}_{Bb,a}- \Gamma^{A}_{Ba,b}
+\Gamma^{A}_{Da}\Gamma^{D}_{Bb} -\Gamma^{A}_{Db}\Gamma^{D}_{Ba}, \nonumber\\
R^{A}_{abB}&=&\Gamma^{A}_{aB,b}
+\Gamma^{A}_{Db}\Gamma^{D}_{aB}
-\Gamma^{A}_{cB}\Gamma^{c}_{ab},\nonumber \\
R^{a}_{bAB}&=&
\Gamma^{a}_{DA}\Gamma^{D}_{bB} -\Gamma^{a}_{EB}\Gamma^{E}_{bA},\nonumber\\
R^{a}_{bcd}&=&\Gamma^{a}_{bd,c}- \Gamma^{a}_{bc,d}
+\Gamma^{a}_{ec}\Gamma^{e}_{bd} -\Gamma^{a}_{ed}\Gamma^{e}_{bc}, \nonumber\\
R^{A}_{BDE}&=&
\Gamma^{A}_{aD}\Gamma^{a}_{BE} -\Gamma^{A}_{aE}\Gamma^{a}_{BD},
\nonumber \\
R^{A}_{BDb}&=&0,\;\;\; R^{A}_{abc}=0. \nonumber
\end{eqnarray}

\noindent
The curvature invariant is defined by

\begin{equation}
I=R^{\mu\nu\alpha\beta}R_{\mu\nu\alpha\beta}.
\end{equation}

\noindent
This can be written as

\begin{equation}
I=R^{abcd}R_{abcd}+R^{ABDE}R_{ABDE}+2R^{ABab}R_{ABab}+4R^{aAbB}R_{aAbB},
\end{equation}
\noindent
where

\begin{eqnarray}
R^{abcd}R_{abcd} &=&R^{2}, \nonumber\\
R^{ABCD}R_{ABCD} &=&{1 \over 8}g^{ab}g^{cd}[tr(\partial_{c}H^{-1}
\partial_{a}H)tr(\partial_{d}H^{-1}\partial_{b}H) \nonumber \\
&-&tr(\partial_{c}H^{-1}\partial_{b}H
\partial_{d}H^{-1}\partial_{a}H)], \nonumber\\
R^{ABab}R_{ABab}&=& {1 \over 8}g^{ab}g^{cd}[
 tr(\partial_{a}H^{-1}\partial_{d}H \partial_{c}H^{-1}\partial_{b}H)
\nonumber \\
&-& tr(\partial_{b}H^{-1}\partial_{c}H \partial_{a}H^{-1}\partial_{d}H)],
 \nonumber  \\
R^{aAbB}R_{aAbB}&=& {1 \over 4}g^{ad}g^{eb}
tr(H^{-1}\bigtriangledown_{a} \bigtriangledown_{b}H H^{-1}
 \bigtriangledown_{d} \bigtriangledown_{e}H) \nonumber \\
 &+&{1 \over 4}g^{ad}g^{eb}
 tr(H^{-1} \bigtriangledown_{a} \bigtriangledown_{b}H H^{-1}\partial_{d}
H H^{-1} \partial_{e}H) \nonumber \\
 &+&{1 \over 16}g^{ad}g^{eb}
 tr(H^{-1} \partial_{b}H H^{-1}\partial_{d}H
H^{-1}\partial_{a}H H^{-1}\partial_{e}H). \nonumber
\end{eqnarray}

\noindent
We may , in general, discuss the singularity structure of colliding 
gravitational
plane waves in $2n+2$ dimensions, but the higher dimensional
Szekeres vacuum solutions give the similar feature of this problem.
First of all the solutions have delta function curvature singularities across
the boundaries $u=0,~v=0$ or $f={1 \over 2}$ and $g={1 \over 2}$ when 
$n_{1}=n_{2}=2$. For other values of $n_{1} >2$ and $n_{2}>2$
curvature has Heaviside step function discontinuity across these boundaries. 
In addition to these discontinuities across the boundaries the
spacetime has essential singularity on the surface $f(u)+g(v)=0$.
For this purpose we shall find the form of the curvature invariant
$I$ as $f+g \rightarrow 0$ which is the singular surface for the four 
dimensional case. We find that

\begin{equation}
I \sim (f_{u}\,g_{v})^2\, (f+g)^{-\mu},
\end{equation}

\noindent
where $\mu=k^2+\ell^2+m^2+2s+2$. For the four dimensional case ($n=1$)
let us choose $k={\bar k}_{1},~\ell={\bar \ell}_{1}$ , $m_{1}=1$, and 
$m^2=1$. Hence in this case
$\mu={\bar k}_{1}^2+{\bar \ell}_{1}^2+2{\bar k}_{1}\, {\bar \ell}_{1}+3$. 
We have both $ 1 \le k^2 <2, ~ 1 \le
\ell^2 <2 $ and $ 1 \le {\bar k}_{1}^2 <2 , ~ 1 \le {\bar \ell}_{1}^2<2$. 
Hence the
constant $m$ plays an important role in the higher dimensional
metrics. On the constants $m_{i}, i=1,2, \cdots, n$ we have the only 
restriction (\ref{em}). Hence as $f+g \rightarrow 0$ we get

$$ {I_{2n+2} \over I_{4}} \sim (f+g)^{1-m^2-2s+2{\bar k}_{1} 
{\bar \ell}_{1}}.$$

\noindent
Here we have made use of conditions (\ref{con}) for $k$ and $\ell$   and  
exactly similar conditions on ${\bar k}_{1}$ and 
${\bar \ell}_{1}$ which implies that
$k^2={\bar k}_{1}^2=2(1- {1 \over n_{1}})$ and $\ell^2={\bar \ell}_{1}^2=
2(1-{1 \over n_{2}})$. 
This means that the singularity structure in the higher dimensional
spacetimes can be made weaker and stronger then the four dimensional
cases by choosing the constants $m_{i}, k_{i}$ and $\ell_{i}$
properly. We have enough freedom to do this for higher values of $n$.

\section{Conclusion}

We have studied the some Ricci flat geometries with arbitrary
signatures. We proved a theorem saying that to all Ricci flat
metrics of four dimensional pseudo Riemannian geometries admitting 
two Killing vector fields there corresponds a class of Ricci flat 
metrics for some $2n+2$ dimensional pseudo Riemannian geometries .
As an application we presented an explicit construction $2n+2$ dimensional 
metrics of colliding gravitational wave spacetimes from a given four
dimensional metrics. We gave a higher dimensional generalization of the
Szekeres metrics and discussed singularity structure of the
corresponding  spacetimes. Further construction of higher dimensional
colliding gravitational plane wave metrics will be communicated
elsewhere \cite{gur3}. A possible extension of our work to low energy
limit of string theory is possible for an arbitrary $n$. Another
application of our approach presented here may be done to the
colliding gravitational plane wave problem for the Einstein-Maxwell-
Dilaton field equations \cite{gur}. 

\vspace{1cm}

This work is partially supported by the Scientific and Technical
Research Council of Turkey (TUBITAK) and by Turkish Academy of Sciences (TUBA).

\end{document}